\documentclass[12pt]{amsart}
\usepackage{amssymb}
\usepackage{epic}
\newtheorem{claim}{}[section]

\newtheorem{proposition}[claim]{Proposition}
\newtheorem{corollary}[claim]{Corollary}
\renewenvironment{proof}{\noindent{\it Proof. \hskip0pt}}
                      {$\square$\par\medskip}

\begin{document}
\baselineskip 6.2 truemm
\parindent 1.5 true pc

\title{Construction of $3\otimes 3$ entangled edge states with positive partial transposes}
\thanks{${}^1$This work was partially supported by KOSEF grant R14-2003-006-01002-0.}
\maketitle

\centerline{by}
\bigskip

\centerline{Kil-Chan Ha} \centerline{Department of Applied
Mathematics, Sejong University, Seoul 143-747, KOREA}
\centerline{\tt kcha@sejong.ac.kr}
\bigskip

\bigskip
\centerline{and}
\bigskip

\centerline{Seung-Hyeok Kye${}^1$} \centerline{Department of
Mathematics, Seoul National University, Seoul 151-742, KOREA}
\centerline{\tt kye@snu.ac.kr}

\vskip 1truecm

\begin{abstract}
We construct a class of $3\otimes 3$ entangled edge states with
positive partial transposes using indecomposable positive linear
maps. This class contains several new types of entangled edge
states with respect to the range dimensions of themselves and
their partial transposes.
\end{abstract}

\newcommand\lan{\langle}
\newcommand\ran{\rangle}
\newcommand\tr{{\text{\rm Tr}}\,}
\newcommand\ot{\otimes}
\newcommand\wt{\widetilde}
\newcommand\join{\vee}
\newcommand\meet{\wedge}
\renewcommand\ker{{\mbox{\rm Ker}}\,}
\newcommand\im{{\mbox{\rm Im}}\,}
\newcommand\mc{\mathcal}
\newcommand\transpose{{\mbox{\rm t}}}
\newcommand\FP{{\mathcal F}({\mathcal P}_n)}
\newcommand\ol{\overline}
\newcommand\JF{{\mathcal J}_{\mathcal F}}
\newcommand\FPtwo{{\mathcal F}({\mathcal P}_2)}
\newcommand\hada{\circledcirc}
\newcommand\id{{\mbox{\rm id}}}
\newcommand\tp{{\mbox{\rm tp}}}
\newcommand\pr{\prime}
\newcommand\inte{{\mbox{\rm int}}\,}
\newcommand\ttt{{\mbox{\rm t}}}
\newcommand\spa{{\mbox{\rm span}}\,}
\newcommand\conv{{\mbox{\rm conv}}\,}
\newcommand\rank{\ {\mbox{\rm rank of}}\ }
\newcommand\vvv{\mathbb V_{m\meet n}\cap\mathbb V^{m\meet n}}
\newcommand\re{{\mbox{\rm Re}}\,}
\newcommand\la{\lambda}
\newcommand\msp{\hskip 2pt}
\newcommand\ppt{\mathbb T}
\newcommand\rk{{\mbox{\rm rank}}\,}


\maketitle

\section{Introduction}

The notion of entanglement in quantum physics has been studied
extensively during the last decade in connection with the quantum
information theory and quantum communication theory. A density
matrix $A$ in $(M_n\otimes M_m)^+$ is said to be {\sl entangled}
if it does not belong to $M_n^+\otimes M_m^+$, where $M_n^+$
denotes the cone of all positive semi-definite $n\times n$
matrices over the complex fields. A density matrix is said to be
{\sl separable} if it belongs to $M_n^+\otimes M_m^+$. Recall that
a density matrix defines a state on the matrix algebra by the
Schur or Hadamard product.

The basic question is, of course, how to distinguish entangled
ones among density matrices, or equivalently among states on
matrices. For a block matrix $A\in M_n\otimes M_m$, the {\sl
partial transpose} or {\sl block transpose} $A^\tau$ of $A$ is
defined by
\[
\left(\sum_{i,j=1}^m a_{ij}\otimes
e_{ij}\right)^\tau=\sum_{i,j=1}^m a_{ji}\otimes e_{ij}.
\]
In the early eighties, it was observed by Choi \cite{choi82} that
the partial transpose of every separable state is positive
semi-definite. This necessary condition for separability has been
also found independently  by Peres \cite{peres}, and is now called the PPT criterion
for separability. Choi \cite{choi82} also gave an example of $3\otimes 3$
entangled state whose partial transpose is positive semi-definite. This kind
of entangled state is called PPTES.

A positive linear map between matrix algebras is said to be {\sl
decomposable} if it is the sum of a completely positive linear map
and a completely copositive linear map. Choi \cite{choi75} was the
first who gave an example of indecomposable positive linear map.
Woronowicz \cite{woronowicz} showed that every positive linear map
from $M_2$ into $M_n$ is decomposable if and only if $n\le 3$. He
showed that there is an indecomposable positive linear map from
$M_2$ into $M_4$ by exhibiting an example of $2\otimes 4$ PPTES.
Str\o rmer \cite{stormer82} also gave a necessary and sufficient
condition for decomposability in terms of partial transpose, and
gave an example of $3\otimes 3$ PPTES. During the nineties, bunch
of examples of PPTES have been found. See
\cite{bdmsst}, \cite{bp}, \cite{dmsst}, \cite{ha+kye}, \cite{hhh},
\cite{p-horo} and \cite{sbl}, for examples. Among examples of
PPTES, so called {\sl edge} PPTES play special roles as was
studied in \cite{lewenstein}.

The cone of all positive semi-definite block matrices with
positive partial transposes will be denoted by $\mathbb T$ in this
paper. The facial structures may be explained in terms of duality
between the space of linear maps and the space of block matrices,
as was studied in \cite{eom-kye} which was motivated by the works
of Woronowicz \cite{woronowicz}, Str\o rmer \cite{stormer82} and
Itoh \cite{itoh}. The cone generated by separable states will be
denoted by $\mathbb V_1$. Then the above mentioned examples will
lie in $\mathbb T\setminus \mathbb V_1$. A PPTES $A$ in $\mathbb
T\setminus \mathbb V_1$ is  an {\sl edge} PPTES if and only if the
proper face of $\mathbb T$ containing $A$ as an interior point
does not contain a separable state.

Edge states may be classified by their range dimensions as was
studied in \cite{sbl}. An edge PPTES $A$ is said to be an $(s,t)$
edge state if the range dimension of $A$ is $s$, and the range
dimension of $A^\tau$ is $t$. Some necessary conditions for
possible combination of $(s,t)$ have been discussed  in
\cite{hlvc} and \cite{sbl}. In the  $3\otimes 3$ cases, it is
quite curious that every known examples of edge PPTES are $(4,4)$
or $(7,6)$ edge states. Here, we assume that $s\ge t$ by the
symmetry. The purpose of this note is to construct another kinds
of $3\otimes 3$ edge states. More precisely, we construct $(7,5)$,
$(6,5)$ and $(8,5)$ edge states as well as $(7,6)$ and $(4,4)$
edge states. It seems to be still open if there exists a $(6,6)$
or $(5,5)$ edge state. This paper was motivated by the paper
\cite{sbl}, where it was conjectured that every $3\otimes 3$
entangled state has Schmidt number two. This is equivalent to ask
if every $2$-positive linear map between $M_3$ is decomposable, by
the duality mentioned above. See \cite{cho-kye-lee}, Corollary 4.3
and \cite{ha} in this direction.

The basic tool is the duality mentioned above. In the Section 2,
we briefly recall the basic notions of the duality, together with
the results in \cite{ha_kye_intersection}, \cite{ha+kye} which
show that every edge state may be constructed from an
indecomposable positive linear map. Our examples of edge states
will be constructed in Section 3 from the indecomposable maps
considered in \cite{cho-kye-lee}.

Throughout this paper, we will not use bra-ket notations. Every vector will be considered as a column vector. If
$x\in\mathbb C^m$ and $y\in \mathbb C^n$ then $x$ will be considered as an $m\times 1$ matrix, and $y^*$ will be
considered as a $1\times n$ matrix, and so $xy^*$ is an $m\times n$ rank one matrix whose range is generated by $x$
and whose kernel is orthogonal to $y$. For a vector $x$, the notation $\ol x$ will
be used for the vector
whose entries are conjugate of the corresponding entries. The notation $\lan\cdot,\cdot \ran$ will
be used for bi-linear pairing. On the other hand, $(\cdot |\cdot )$ will be used for the inner product, which is
sesqui-linear, that is, linear in the first variable and conjugate-linear in the second variable.
For natural numbers $m$ and $n$, we denote by $m\join n$ and $m\meet n$ the maximum and minimum of $m$ and $n$, respectively.

\section{Decomposable maps and PPT entanglement}

For a given finite set ${\mathcal V}=\{V_1, V_2,\dots, V_\nu\}\subset M_{m\times n}$ of $m\times n$ matrices, we define
linear maps $\phi_\mathcal V$ and $\phi^\mathcal V$ from $M_m$ into $M_n$ by the following:
\[
\begin{aligned} &\phi_\mathcal V: X\mapsto \sum_{i=1}^\nu
V_i^*XV_i,\qquad X\in
M_m,\\
&\phi^\mathcal V: X\mapsto \sum_{i=1}^\nu V_i^*X^{\ttt}V_i,\qquad
X\in M_m,
\end{aligned}
\]
where $X^\ttt$ denotes   the transpose of $X$. We denote by
$\phi_V=\phi_{\{V\}}$ and $\phi^V=\phi^{\{V\}}$. It is well-known
\cite{choi75-10}, \cite{kraus} that every completely positive
(respectively completely copositive) linear map between matrix
algebras is of the form $\Phi_\mathcal V$ (respectively $\Phi^\mathcal V$). We denote by $\mathbb P_{m\meet n}$
(respectively $\mathbb P^{m\meet n}$) the convex cone of all
completely positive (respectively completely copositive) linear
maps. For a subspace $E$ of $M_{m\times n}$, we define
\[
\begin{aligned} \Phi_E=&\{\phi_\mathcal V\in\mathbb P_{m\meet n}:
\spa{\mathcal
V}\subset E\},\\
\Phi^E=&\{\phi^\mathcal V\in\mathbb P^{m\meet n}: \spa{\mathcal
V}\subset E\},
\end{aligned}
\]
where $\spa{\mathcal V}$ denotes the span of the set $\mathcal V$.
We have shown in \cite{kye-cambridge} that the correspondence
\[
E  \mapsto \Phi_E\qquad {\mbox{\rm (respectively}}\ E\mapsto
\Phi^E{\mbox{\rm )}}
\]
gives rise to a lattice isomorphism between the lattice
of all subspaces of the vector space $M_{m\times n}$ and the lattice
of all faces of the convex cones $\mathbb P_{m\meet n}$ (respectively $\mathbb P^{m\meet n}$).
A linear map in the cone
\[
\mathbb D:= \conv (\mathbb P_{m\meet n}, \mathbb P^{m\meet  n})
\]
is said to be {\sl decomposable}, where $\conv (C_1,C_2)$ denotes the
convex hull of $C_1$ and $C_2$. Every decomposable map is
positive, that is, sends positive semi-definite matrices into
themselves, but the converse is not true. There are many examples
of indecomposable positive linear maps in the literature
\cite{cho-kye-lee}, \cite{choi75}, \cite{ha}, \cite{ha-1}, \cite{ha-2},
\cite{kim-kye}, \cite{kye-22}, \cite{osaka}, \cite{robertson},
\cite{stormer82}, \cite{tomiyama}, \cite{tang}, \cite{terhal}. We
have shown in \cite{kye-decom} that every face of the cone
$\mathbb D$ is of the form
\[
\sigma (D,E):=\conv (\Phi_D,\Phi^E)
\]
for a pair
$(D,E)$ of subspaces of $M_{m\times n}$. This pair of subspaces is uniquely determined under the assumption
\begin{equation}\label{cond-decom}
\sigma (D,E)\cap \mathbb P_{m\meet n}=\Phi_D,\qquad
\sigma (D,E)\cap \mathbb P^{m\meet n}=\Phi^E.
\end{equation}
We say that a pair
$(D,E)$ is a {\sl decomposition pair} if the set $\conv
(\Phi_D,\Phi^E)$ is a face of $\mathbb D$ with the condition (\ref{cond-decom}). Faces of the cone $\mathbb D$ and
decomposition pairs correspond each other  in this way.
 Whenever we use the notation $\sigma(D,E)$, we assume that
$(D,E)$ is a decomposition pair. It is very hard to determine all decomposition pairs. See
\cite{byeon-kye}, \cite{kye-2by2_II} for the simplest case of $m=n=2$.

Now, we turn our attention to the block matrices, and identify an $m\times n$ matrix $z\in M_{m\times n}$  and a
vector $\wt z\in \mathbb C^n\otimes\mathbb C^m$ as follows: For
$z=[z_{ik}]\in M_{m\times n}$, define
\[
\begin{aligned}
z_i=&\sum_{k=1}^n z_{ik} e_k\in \mathbb C^n,\qquad i=1,2,\dots, m,\\
\wt z=&\sum_{i=1}^m z_i\otimes e_i\in \mathbb C^n\otimes\mathbb
C^m.
\end{aligned}
\]
Then $z\mapsto \wt z$ defines an inner product isomorphism from
$M_{m\times n}$ onto $\mathbb C^n\otimes \mathbb C^m$. We also
note that $\wt z\,{\wt z}^*$ is a positive semi-definite matrix in
$M_n\otimes M_m$ of rank one.
We consider the convex cones
\[
\begin{aligned}
\mathbb V_s&=\conv\{\wt z\,\wt z^*: {\mbox{\rm rank}}\ z\le s\},\\
\mathbb V^s&=\conv\{(\wt z\,\wt z^*)^\tau: {\mbox{\rm rank}}\ z\le
s\}.
\end{aligned}
\]
for $s=1,2,\dots, m\meet n$.
By the relation
\[
\wt{xy^*}\,{\wt{xy^*}}^*=(\ol y\otimes x)(\ol y\otimes x)^*
=\ol y\,{\ol y}^*\otimes xx^*,\qquad x\in\mathbb C^m,\ y\in\mathbb C^n,
\]
we have
\[
\mathbb V_1=M_n^+\otimes M_m^+.
\]
Therefore, a density matrix in $M_n\otimes M_m$ is
separable if and only if it belongs to
the cone $\mathbb V_1$.
If $z=xy^*$ is a rank one matrix
with column vectors $x\in\mathbb C^m, y\in\mathbb C^n$
then $(zz^*)^\tau=ww^*$ is positive semi-definite with $w=\ol xy^*$
by  a direct simple calculation.
Therefore, we see \cite{choi82}, \cite{peres} that every separable state belongs to the convex cone
\[
\mathbb T:=\mathbb V_{m\meet n}\cap\mathbb V^{m\meet n}=\{A\in (M_n\otimes M_m)^+: A^\tau\in(M_n\otimes M_m)^+\}.
\]
A block matrix in the cone $\mathbb T$ is said to be {\sl of positive partial transpose}.

It is well known that every face of $\mathbb V_{m\meet n}=(M_n\otimes M_m)^+$ and $\mathbb V^{m\meet n}$ is of the form
\[
\begin{aligned}
\Psi_D&=\{A\in (M_n\otimes M_m)^+: {\mathcal R}A\subset \wt D\},\\
\Psi^E&=\{A\in M_n\otimes M_m:  A^\tau\in\Psi_E\},
\end{aligned}
\]
respectively, where ${\mathcal R}A$ is the range space of $A$ and $\wt D=\{\wt z\in\mathbb C^n\otimes\mathbb C^m: z\in D\}$.
It is also easy to see that
every face of $\mathbb T$ is of the form
 \[
\tau (D,E):=\Psi_D \cap \Psi^E
\]
for a pair $(D,E)$ of subspaces of $M_{m\times n}$, as was explained in
\cite{ha_kye_intersection}. This pair is uniquely determined under the assumption
\begin{equation}\label{cond-inter}
\inte \tau (D,E)\subset \inte \Psi_D,\qquad
\inte \tau  (D,E)\subset \inte \Psi^E,
\end{equation}
where $\inte C$ denote the relative interior of the convex set $C$ with respect to
hyperplane generated by $C$. We say that a pair $(D,E)$ of subspaces is an {\sl intersection pair} if
it satisfies the assumption (\ref{cond-inter}), and $\tau  (D,E)\neq \emptyset$.
We also assume the condition (\ref{cond-inter}) whenever we use the notation
$\tau(D,E)$.

Note that the convex cone $\mathbb D$ and $\mathbb T$ are sitting in the vector space
${\mathcal L}(M_m,M_n)$ of all linear maps from $M_m$ into $M_n$ and the vector space
$M_n\otimes M_m$ of all block matrices.
In \cite{eom-kye}, we have considered the bi-linear pairing between
the spaces ${\mathcal L}(M_m, M_n)$ and $M_n\otimes M_m$, given by
\begin{equation}\label{definition}
\lan A,\phi\ran =\tr \left[ \left(\sum_{i,j=1}^m \phi(e_{ij})
\otimes e_{ij}\right) A^\ttt \right]
=\sum_{i,j=1}^m\lan\phi(e_{ij}),a_{ij}\ran,
\end{equation}
for $A=\sum_{i,j=1}^m a_{ij}\ot e_{ij}\in M_n\ot M_m$ and
$\phi\in{\mathcal L}(M_m, M_n)$, where the bi-linear form in the
right-hand side is given by $\lan X,Y\ran=\tr (YX^\ttt )$ for
$X,Y\in M_n$. The main result of \cite{eom-kye} tells us that two
cones $\mathbb D$ and $\mathbb T$ are dual each other in the
following sense:
\[
\begin{aligned}
\phi\in\mathbb D\ &\Longleftrightarrow\ \lan A,\phi\ran\ge 0\ {\mbox{\rm for every}}\ A\in\mathbb T,\\
A\in\mathbb T\ &\Longleftrightarrow\ \lan A,\phi\ran\ge 0\
{\mbox{\rm for every}}\ \phi\in\mathbb D.
\end{aligned}
\]
It was also shown in \cite{eom-kye} that the cone $\mathbb P_s$ (respectively $\mathbb P^s$)
consisting of $s$-positive (respectively $s$-copositive) linear maps is dual to the cone
$\mathbb V_s$ (respectively $\mathbb V ^s$) in the above sense. See also \cite{sbl}.
Some faces of the cone $\mathbb D$ arise from this duality, and is of the form
\[
\tau(D,E)^\prime:= \{\phi\in \mathbb D: \lan A,\phi\ran =0\
{\mbox{\rm for every}}\ A\in\tau(D,E)\}
\]
for a face $\tau(D,E)$ of $\mathbb T$. If $A$ is an interior point of $\tau(D,E)$ then we have
\[
\tau(D,E)^\prime=A^\prime:=\{\phi\in \mathbb D: \lan A,\phi\ran =0\}.
\]
It is easy to see that
\[
\tau(D,E)^\prime=\sigma(D^\perp, E^\perp).
\]
It should be noted that not every face arises in this way even in the simplest case of
$m=n=2$. See \cite{byeon-kye}, \cite{kye-2by2_II}. Nevertheless, every face of the cone $\mathbb T$ arises from this duality.
More precisely, it was shown in \cite{ha_kye_intersection} that every face of the cone $\mathbb T$ is of the form
\[
\sigma(D,E)^\prime:= \{A\in \mathbb T: \lan A,\phi\ran =0\
{\mbox{\rm for every}}\ \phi\in\sigma(D,E)\}=\tau(D^\perp,
E^\perp)
\]
for a face $\sigma(D,E)$ of the cone $\mathbb D$. The following is implicit in \cite{ha_kye_intersection}. We state here
for the clearance.

\begin{proposition}
A pair $(D,E)$ of subspaces of $M_{m\times n}$ is an intersection pair if and only if there exists
$A\in\mathbb T$ such that ${\mathcal R}A=\wt D$ and ${\mathcal R}A^\tau=\wt E$.
If this is the case then we have
\[
\inte\tau(D,E)=\{A\in\mathbb T: {\mathcal R}A=\wt D,\ {\mathcal R}A^\tau=\wt E\}.
\]
\end{proposition}

\begin{proof}
Let $(D,E)$ be an intersection pair and take $A\in\inte\tau(D,E)$. Then
$A^\prime=\tau(D,E)^\prime=\sigma(D^\perp, E^\perp)$, and we have
${\mathcal R}A=\wt D$ and ${\mathcal R}A^\tau=\wt E$ by \cite{ha_kye_intersection} Lemma 1.
For the converse, assume that there is $A\in\mathbb T$ such that
${\mathcal R}A=\wt D$ and ${\mathcal R}A^\tau=\wt E$. Take the intersection pair $(D_1,E_1)$ such that
$A\in\inte\tau(D_1,E_1)$ Then we have
${\mathcal R}A=\wt D_1$ and ${\mathcal R}A^\tau=\wt E_1$, and so $D=D_1$ and $E=E_1$.
The last statement has been already proved.
\end{proof}

\begin{corollary}
If $(D_1,E_1)$ and $(D_2,E_2)$ are intersection pairs then $(D_1\join D_2, E_1\join E_2)$ is also an intersection
pair.
\end{corollary}

\begin{proof}
Take $A_i\in\mathbb T$ with ${\mathcal R}A_i=\wt D_i$ and ${\mathcal R}A_i^\tau=\wt E_i$ for $i=1,2$. Then we have
\[
A_1+A_2\in\mathbb T,\qquad {\mathcal R}(A_1+A_2)=\wt {D_1\join D_2},\qquad {\mathcal R}(A_1+A_2)^\tau=\wt {E_1\join E_2}.
\]
Therefore, we see that $(D_1\join D_2, E_1\join E_2)$ is an intersection
pair.
\end{proof}

Now, we have two cones $\mathbb D\subset\mathbb P_1$ in the space ${\mathcal L}(M_m,M_n)$ and another two cones
$\mathbb V_1\subset\mathbb T$ in the space $M_n\otimes M_m$. Recall that $\mathbb P_1$ denotes the cone of all
positive linear maps. The pairs $(\mathbb D,\mathbb T)$ and $(\mathbb P_1,\mathbb V_1)$ are dual each other, as was explained before.
Let $\sigma(D,E)$ be a proper face of the cone $\mathbb D$. Then we have the following two cases:
\[
\inte\sigma(D,E)\subset\inte\mathbb P_1\qquad {\mbox{\rm
or}}\qquad \sigma(D,E)\subset\partial\mathbb P_1,
\]
since $\sigma(D,E)$ is a convex subset of the cone $\mathbb P_1$, where
$\partial C:=C\setminus \inte C$ denotes the boundary of the convex set $C$. We have shown in
\cite{ha_kye_intersection}, \cite{ha+kye}
that
\begin{equation}\label{edge_dual}
\inte\sigma(D,E)\subset \inte\mathbb P_1\ \Longleftrightarrow\
\sigma(D,E)^\prime\cap \mathbb V_1=\{0\}.
\end{equation}
Every element $A\in\mathbb T$ determines a unique face $\tau(D,E)$ whose interior contains $A$.
Then a density block matrix
$A\in\mathbb T$ is an
(entangled) edge state if and only if $\tau(D,E)\cap\mathbb V_1=\{0\}$. Therefore, we conclude the following:
\begin{enumerate}
\item
If $\sigma(D,E)$ is a face of $\mathbb D$ with $\inte\sigma(D,E)\subset\inte\mathbb P_1$ then every nonzero element in
the dual face $\sigma(D,E)^\prime$ gives rise to an entangled edge state up to constant multiplications,
\item
Every edge state arises in this way.
\end{enumerate}
The second claim follows from the fact that every face of the cone $\mathbb T$ arises from the duality,
as was explained before.

\section{Construction of $3\otimes 3$ PPT entangled edge states}

We  begin with the decomposable positive linear map $\phi:M_3\to M_3$ defined by
\[
\phi=\phi_{e_{11}-e_{22}}+\phi_{e_{22}-e_{33}}+\phi_{e_{33}-e_{11}}+
\phi^{\mu e_{12}-\la e_{21}}+\phi^{\mu e_{23}-\la
e_{32}}+\phi^{\mu e_{31}-\la e_{13}},
\]
which lies in $\partial\mathbb D\cap\inte\mathbb P_1$ as was shown
in \cite{ha+kye}, where
\[
\la\mu=1,\qquad \lambda>0,\qquad \lambda\neq 1.
\]
We try to determine the dual face
$\tau(D,E)=\{\phi\}^{\prime}$. This map was originated from
indecomposable positive linear maps considered in
\cite{cho-kye-lee}.
We note that $D$ is the
$7$-dimensional space given by
\[
D=\spa\{e_{12},\ e_{21},\ e_{23},\  e_{32},\ e_{31},\ e_{13},\ e_{11}+e_{22}+e_{33}\},
\]
and $E$ is the $6$-dimensional space given by
\[
E=\spa\{\la e_{12}+\mu e_{21},\ \la e_{23}+\mu e_{32},\ \la e_{31}+\mu e_{13},\ e_{11},\ e_{22},\ e_{33}\}.
\]
Therefore, every matrix $x_i\in E$ is of the form
\[
x_i=\rho\circ\sigma_i
\]
where
\[
\rho= \left(
\begin{array}{ccc}
1&\la&\mu\\ \mu&1&\la\\ \la&\mu &1
\end{array}\right),
\quad \sigma_i= \left(
\begin{array}{ccc}
\xi_i&\alpha_i&\gamma_i\\ \alpha_i&\eta_i&\beta_i\\
\gamma_i&\beta_i&\zeta_i
\end{array}\right),
\]
and $\rho \circ\sigma_i$ denotes the Hadamard product of $\rho$
and $\sigma_i$.

 It follows that if $X^\tau=\sum_i\wt x_i\wt x_i^*\in\mathbb
V_3\cap\mathbb V^3$ belongs to $\tau(D,E)$ then
\[
X^\tau=\sum (\wt\rho \wt\rho^*)\circ(\wt\sigma_i \wt\sigma_i^*)
=(\wt\rho \wt\rho^*)\circ Y
\]
with
\[
\wt\rho\wt\rho^*= \left(
\begin{array}{ccccccccccc}
1     &\la   &\mu  &&\mu  &1     &\la   &&\la   &\mu  &1     \\
\la   &\la^2 &1    &&1    &\la   &\la^2 &&\la^2 &1    &\la   \\
\mu  &1    &\mu^2 &&\mu^2 &\mu  &1    &&1    &\mu^2 &\mu  \\
\\
\mu  &1    &\mu^2 &&\mu^2 &\mu  &1    &&1    &\mu^2 &\mu  \\
1     &\la   &\mu  &&\mu  &1     &\la   &&\la   &\mu  &1     \\
\la   &\la^2 &1    &&1    &\la   &\la^2 &&\la^2 &1    &\la   \\
\\
\la   &\la^2 &1    &&1    &\la   &\la^2 &&\la^2 &1    &\la   \\
\mu  &1    &\mu^2 &&\mu^2 &\mu  &1    &&1    &\mu^2 &\mu  \\
1     &\la   &\mu  &&\mu  &1     &\la   &&\la   &\mu  &1
\end{array}
\right),
\]
and $Y= \sum \wt\sigma_i\wt\sigma_i^*$ is given by
\[
\left(
\begin{array}{ccccccccccc}
(\xi|\xi) &(\xi|\alpha)  &(\xi|\gamma) &&(\xi|\alpha)  &(\xi|\eta) &(\xi|\beta) &&(\xi|\gamma) &(\xi|\beta) &(\xi|\zeta) \\
(\alpha|\xi)  &(\alpha|\alpha)   &(\alpha|\gamma)  &&(\alpha|\alpha)   &(\alpha|\eta)  &(\alpha|\beta)  &&(\alpha|\gamma)  &(\alpha|\beta)  &(\alpha|\zeta)  \\
(\gamma|\xi) &(\gamma|\alpha)  &(\gamma|\gamma) &&(\gamma|\alpha)  &(\gamma|\eta) &(\gamma|\beta) &&(\gamma|\gamma) &(\gamma|\beta) &(\gamma|\zeta) \\
\\
(\alpha|\xi)  &(\alpha|\alpha)   &(\alpha|\gamma)  &&(\alpha|\alpha)   &(\alpha|\eta)  &(\alpha|\beta)  &&(\alpha|\gamma)  &(\alpha|\beta)  &(\alpha|\zeta)  \\
(\eta|\xi) &(\eta|\alpha)  &(\eta|\gamma) &&(\eta|\alpha)  &(\eta|\eta) &(\eta|\beta) &&(\eta|\gamma) &(\eta|\beta) &(\eta|\zeta) \\
(\beta|\xi) &(\beta|\alpha)  &(\beta|\gamma) &&(\beta|\alpha)  &(\beta|\eta) &(\beta|\beta) &&(\beta|\gamma) &(\beta|\beta) &(\beta|\zeta) \\
\\
(\gamma|\xi) &(\gamma|\alpha)  &(\gamma|\gamma) &&(\gamma|\alpha)  &(\gamma|\eta) &(\gamma|\beta) &&(\gamma|\gamma) &(\gamma|\beta) &(\gamma|\zeta) \\
(\beta|\xi) &(\beta|\alpha)  &(\beta|\gamma) &&(\beta|\alpha)  &(\beta|\eta) &(\beta|\beta) &&(\beta|\gamma) &(\beta|\beta) &(\beta|\zeta) \\
(\zeta|\xi) &(\zeta|\alpha)  &(\zeta|\gamma) &&(\zeta|\alpha)  &(\zeta|\eta) &(\zeta|\beta) &&(\zeta|\gamma) &(\zeta|\beta) &(\zeta|\zeta) \\
\end{array}
\right)
\]
if we denote by $\xi,\eta, \zeta, \alpha,\beta$ and $\gamma$ the
vectors whose entries are $\xi_i,\eta_i, \zeta_i,
\alpha_i,\beta_i$ and $\gamma_i$, respectively. So,
$X=(X^\tau)^\tau$ is of the form
\[
\left(
\begin{array}{ccccccccccc}
(\xi|\xi)      &\la(\xi|\alpha)      &\mu(\xi|\gamma)       &&\mu(\alpha|\xi)   &(\alpha|\alpha)   &\mu^2(\alpha|\gamma)  &&\la(\gamma|\xi)   &\la^2(\gamma|\alpha)   &(\gamma|\gamma)\\
\la(\alpha|\xi)   &\la^2(\alpha|\alpha)   &(\alpha|\gamma)  &&(\eta|\xi)     &\la(\eta|\alpha)     &\mu(\eta|\gamma)    &&\mu(\beta|\xi)    &(\beta|\alpha)    &\mu^2(\beta|\gamma)   \\
\mu(\gamma|\xi)   &(\gamma|\alpha)   &\mu^2(\gamma|\gamma)  &&\la(\beta|\xi)    &\la^2(\beta|\alpha)    &(\beta|\gamma)   &&(\zeta|\xi)    &\la(\zeta|\alpha)      &\mu(\zeta|\gamma) \\
\\
\mu(\xi|\alpha)  &(\xi|\eta) &\la(\xi|\beta)               &&\mu^2(\alpha|\alpha)   &\mu(\alpha|\eta)  &(\alpha|\beta)    &&(\gamma|\alpha)  &\la(\gamma|\eta) &\la^2(\gamma|\beta)  \\
(\alpha|\alpha)   &\la(\alpha|\eta)  &\la^2(\alpha|\beta)     &&\mu(\eta|\alpha)  &(\eta|\eta) &\la(\eta|\beta)            &&\mu^2(\beta|\alpha)  &\mu(\beta|\eta) &(\beta|\beta) \\
\mu^2(\gamma|\alpha)  &\mu(\gamma|\eta) &(\gamma|\beta)       &&(\beta|\alpha)  &\la(\beta|\eta) &\la^2(\beta|\beta)          &&\mu(\zeta|\alpha)  &(\zeta|\eta) &\la(\zeta|\beta) \\
\\
\la(\xi|\gamma) &\mu(\xi|\beta) &(\xi|\zeta)               &&(\alpha|\gamma)  &\mu^2(\alpha|\beta)  &\mu(\alpha|\zeta)   &&\la^2(\gamma|\gamma) &(\gamma|\beta) &\la(\gamma|\zeta) \\
\la^2(\alpha|\gamma)  &(\alpha|\beta)  &\la(\alpha|\zeta)   &&\la(\eta|\gamma) &\mu(\eta|\beta) &(\eta|\zeta)         &&(\beta|\gamma) &\mu^2(\beta|\beta) &\mu(\beta|\zeta) \\
(\gamma|\gamma) &\mu^2(\gamma|\beta) &\mu(\gamma|\zeta)    &&\la^2(\beta|\gamma) &(\beta|\beta) &\la(\beta|\zeta)           &&\la(\zeta|\gamma) &\mu(\zeta|\beta) &(\zeta|\zeta) \\
\end{array}
\right)
\]

Now, we consider the condition $X\in (\Phi_{D^\perp})^\prime$ to see that
\[
\lan Y,\wt\rho\wt\rho^*\circ\phi^V\ran
=\lan \wt\rho\wt\rho^*\circ Y,\phi^V\ran
=\lan X^\tau,\phi^V\ran
=\lan X,\phi_V\ran=0
\]
for any $V\in D^\perp$. Note that any matrix $V$ in $D^\perp$ is of the form
\[
V=\left(
\begin{array}{ccc}
a_1&\cdot&\cdot\\
\cdot&a_2&\cdot\\
\cdot&\cdot&a_3
\end{array}\right),\qquad a_1+a_2+a_3=0,
\]
and
\[
\wt\rho\wt\rho^*\circ\phi^V=
\left(
\begin{array}{ccccccccccc}
|a_1|^2     & \cdot &\cdot  &&\cdot  &\cdot     &\cdot  &&\cdot  &\cdot  &\cdot     \\
\cdot&\cdot     &\cdot  &&\ol a_2a_1 &\cdot  &\cdot  &&\cdot  &\cdot  &\cdot  \\
\cdot&\cdot     &\cdot  &&\cdot &\cdot  &\cdot  &&\ol a_3a_1  &\cdot  &\cdot  \\
\\
\cdot&\ol a_1a_2     &\cdot  &&\cdot &\cdot  &\cdot  &&\cdot  &\cdot  &\cdot  \\
\cdot     &\cdot  &\cdot  &&\cdot  &|a_2|^2     &\cdot  &&\cdot  &\cdot  &\\
\cdot&\cdot     &\cdot  &&\cdot &\cdot  &\cdot  &&\cdot  &\ol a_3a_2  &\cdot  \\
\\
\cdot&\cdot     &\ol a_1a_3  &&\cdot &\cdot  &\cdot  &&\cdot  &\cdot  &\cdot  \\
\cdot&\cdot     &\cdot  &&\cdot &\cdot  &\ol a_2a_3  &&\cdot  &\cdot  &\cdot  \\
\cdot     &\cdot  &\cdot  &&\cdot  &\cdot     &\cdot  &&\cdot  &\cdot  &|a_3|^2
\end{array}
\right).
\]
Therefore, it follows that
\[
\left(
\begin{array}{ccc}
(\xi|\xi) &(\alpha|\alpha)  &(\gamma|\gamma)\\
(\alpha|\alpha) &(\eta|\eta) &(\beta|\beta)\\
(\gamma|\gamma)&(\beta|\beta)&(\zeta|\zeta)
\end{array}\right)\circ
\left(
\begin{array}{ccc}
|a_1|^2     & \ol a_2a_1 &\ol a_3a_1\\
\ol a_1a_2     &|a_2|^2  &\ol a_3a_2\\
\ol a_1a_3  &\ol a_2a_3&|a_3|^2
\end{array}
\right)=0
\]
whenever $a_1+a_2+a_3=0$, where $A\circ B=\sum a_{ij}b_{ij}$ by an abuse of notation.
Taking $(a_1,a_2,a_3)=(1,-1,0)$, we have
$\|\xi\|^2+\|\eta\|^2=2\|\alpha\|^2$. But the positivity of the
$2\times 2$ submatrix of $X$ with the 1, 5 columns and rows tells
us that $\|\xi\|=\|\eta\|=\|\alpha\|$. Similarly, we have
\[
\|\xi\|=\|\eta\|=\|\zeta\|=\|\alpha\|=\|\beta\|=\|\gamma\|=1
\]
by assuming that $\|\xi\|=1$.
Hence, $X$ is of the form
\[
\left(
\begin{array}{ccccccccccc}
1      &\la(\xi|\alpha)      &\mu(\xi|\gamma)       &&\mu(\alpha|\xi)   &1   &\mu^2(\alpha|\gamma)  &&\la(\gamma|\xi)   &\la^2(\gamma|\alpha)   &1\\
\la(\alpha|\xi)   &\la^2   &(\alpha|\gamma)  &&(\eta|\xi)     &\la(\eta|\alpha)     &\mu(\eta|\gamma)    &&\mu(\beta|\xi)    &(\beta|\alpha)    &\mu^2(\beta|\gamma)   \\
\mu(\gamma|\xi)   &(\gamma|\alpha)   &\mu^2  &&\la(\beta|\xi)    &\la^2(\beta|\alpha)    &(\beta|\gamma)   &&(\zeta|\xi)    &\la(\zeta|\alpha)      &\mu(\zeta|\gamma) \\
\\
\mu(\xi|\alpha)  &(\xi|\eta) &\la(\xi|\beta)               &&\mu^2   &\mu(\alpha|\eta)  &(\alpha|\beta)    &&(\gamma|\alpha)  &\la(\gamma|\eta) &\la^2(\gamma|\beta)  \\
1   &\la(\alpha|\eta)  &\la^2(\alpha|\beta)     &&\mu(\eta|\alpha)  &1 &\la(\eta|\beta)            &&\mu^2(\beta|\alpha)  &\mu(\beta|\eta) &1 \\
\mu^2(\gamma|\alpha)  &\mu(\gamma|\eta) &(\gamma|\beta)       &&(\beta|\alpha)  &\la(\beta|\eta) &\la^2          &&\mu(\zeta|\alpha)  &(\zeta|\eta) &\la(\zeta|\beta) \\
\\
\la(\xi|\gamma) &\mu(\xi|\beta) &(\xi|\zeta)               &&(\alpha|\gamma)  &\mu^2(\alpha|\beta)  &\mu(\alpha|\zeta)   &&\la^2 &(\gamma|\beta) &\la(\gamma|\zeta) \\
\la^2(\alpha|\gamma)  &(\alpha|\beta)  &\la(\alpha|\zeta)   &&\la(\eta|\gamma) &\mu(\eta|\beta) &(\eta|\zeta)         &&(\beta|\gamma) &\mu^2 &\mu(\beta|\zeta) \\
1 &\mu^2(\gamma|\beta) &\mu(\gamma|\zeta)    &&\la^2(\beta|\gamma) &1 &\la(\beta|\zeta)           &&\la(\zeta|\gamma) &\mu(\zeta|\beta) &1 \\
\end{array}
\right)
\]
If we take vectors so that
$\spa\{\xi,\eta,\zeta\}\perp\spa\{\alpha,\beta,\gamma\}$ with
mutually orthonormal vectors $\alpha,\beta,\gamma$ then we have
\begin{equation}\label{eq:xx}
X= \left(
\begin{array}{ccccccccccc}
1      &\cdot      &\cdot       &&\cdot   &1   &\cdot  &&\cdot   &\cdot   &1\\
\cdot   &\la^2   &\cdot  &&(\eta|\xi)     &\cdot     &\cdot    &&\cdot    &\cdot    &\cdot   \\
\cdot   &\cdot   &\mu^2  &&\cdot    &\cdot    &\cdot   &&(\zeta|\xi)    &\cdot      &\cdot \\
\\
\cdot  &(\xi|\eta) &\cdot               &&\mu^2   &\cdot  &\cdot    &&\cdot  &\cdot &\cdot  \\
1   &\cdot  &\cdot     &&\cdot  &1 &\cdot            &&\cdot  &\cdot &1 \\
\cdot  &\cdot &\cdot      &&\cdot  &\cdot &\la^2          &&\cdot  &(\zeta|\eta) &\cdot \\
\\
\cdot &\cdot &(\xi|\zeta)               &&\cdot  &\cdot  &\cdot   &&\la^2 &\cdot &\cdot \\
\cdot  &\cdot  &\cdot  &&\cdot &\cdot &(\eta|\zeta)         &&\cdot &\mu^2&\cdot \\
1 &\cdot &\cdot    &&\cdot &1 &\cdot           &&\cdot &\cdot &1 \\
\end{array}
\right)
\end{equation}
and
\[
X^\tau=
\left(
\begin{array}{ccccccccccc}
1      &\cdot      &\cdot       &&\cdot   &(\xi|\eta)   &\cdot  &&\cdot   &\cdot   &(\xi|\zeta)\\
\cdot   &\la^2   &\cdot  &&1     &\cdot     &\cdot    &&\cdot    &\cdot    &\cdot   \\
\cdot   &\cdot   &\mu^2  &&\cdot    &\cdot    &\cdot   &&1    &\cdot      &\cdot \\
\\
\cdot  &1 &\cdot               &&\mu^2   &\cdot  &\cdot    &&\cdot  &\cdot &\cdot  \\
(\eta|\xi)   &\cdot  &\cdot     &&\cdot  &1 &\cdot            &&\cdot  &\cdot &(\eta|\zeta) \\
\cdot  &\cdot &\cdot      &&\cdot  &\cdot &\la^2          &&\cdot  &1 &\cdot \\
\\
\cdot &\cdot &1               &&\cdot  &\cdot  &\cdot   &&\la^2 &\cdot &\cdot \\
\cdot  &\cdot  &\cdot  &&\cdot &\cdot &1         &&\cdot &\mu^2&\cdot \\
(\zeta|\xi) &\cdot &\cdot    &&\cdot &(\zeta|\eta) &\cdot           &&\cdot &\cdot &1 \\
\end{array}
\right).
\]

We note that the rank of $X$ is equal to
\[
1+\rk\left(\begin{array}{cc}(\xi|\xi)&(\xi|\eta)\\(\eta|\xi)&(\eta|\eta)\end{array}\right)
+\rk\left(\begin{array}{cc}(\eta|\eta)&(\eta|\zeta)\\(\zeta|\eta)&(\zeta|\zeta)\end{array}\right)
+\rk\left(\begin{array}{cc}(\zeta|\zeta)&(\zeta|\xi)\\(\xi|\zeta)&(\xi|\xi)\end{array}\right)
\]
and the rank of $X^\tau$ is equal to
\[
3+\rk\left(\begin{array}{ccc}
(\xi|\xi) & (\xi|\eta) & (\xi|\zeta)\\
(\eta|\xi) &(\eta|\eta)& (\eta|\zeta)\\
(\zeta|\xi) &(\zeta|\eta)&(\zeta|\zeta)
\end{array}\right).
\]
Recall that the rank of the $n\times n$ matrix
$[(\xi_i|\xi_j)]_{i,j=1}^n$ is the dimension of the space
$\spa\{\xi_1,\dots,\xi_n\}$. If we take mutually independent
vectors $\xi,\eta,\zeta$ then we get a $(7,6)$ edge state. If we
take vectors so that $\dim\spa\{\xi,\eta,\zeta\}=2$ and none of
two vectors are linearly dependent then we may get a $(7,5)$ edge
state. If we take vectors so that $\dim\spa\{\xi,\eta,\zeta\}=2$
and one pair of two vectors are linearly dependent then we have a
$(6,5)$ edge state. Finally, if we take vectors with
$\xi=\eta=\zeta$ then we have a $(4,4)$ edge state as was given in
the paper \cite{ha+kye}.
For more explicit examples, we put
$$
\xi=e_1,\qquad \eta=e_2
$$
in $\mathbb C^3$. We get one parameter family of $(7,6)$ edge states
(respectively $(7,5)$ and $(6,5)$ edge states) if we put
$$
\zeta=e_3\qquad {\rm (respectively}\quad \zeta=\frac 1{\sqrt
2}\,(e_1+e_2) \quad {\rm and}\quad \zeta=e_1)
$$
in the matrix (\ref{eq:xx}).

In order to get another edge states such as $(8,5)$ edge states,
we discard the condition $X\in (\Phi_{D^\perp})^\prime$. We define
vectors $\xi,\eta,\zeta\in \mathbb C^5$ by
\[\xi=\sqrt t\, e_1, \quad \eta=\sqrt t\, e_2, \quad
\zeta=\sqrt{\frac 1{t(t+1)}}\,(\xi+\eta)
\] for $t>1$. We also take mutually orthonormal vectors $\alpha,
\beta, \gamma \in \mathbb C^5$ in (\ref{eq:xx}) so that
\[\spa\{\xi,\eta,\zeta\}\perp \spa\{\alpha,\beta,\gamma\}.
\]
Then we have
\[
X= \left(
\begin{array}{ccccccccccc}
t     &\cdot      &\cdot       &&\cdot   &1   &\cdot  &&\cdot   &\cdot   &1\\
\cdot   &\la^2   &\cdot  &&\cdot     &\cdot     &\cdot    &&\cdot    &\cdot    &\cdot   \\
\cdot   &\cdot   &\mu^2  &&\cdot    &\cdot    &\cdot   &&\sqrt{\frac{t}{t+1}}    &\cdot      &\cdot \\
\\
\cdot  &\cdot &\cdot               &&\mu^2   &\cdot  &\cdot    &&\cdot  &\cdot &\cdot  \\
1   &\cdot  &\cdot     &&\cdot  &t &\cdot            &&\cdot  &\cdot &1 \\
\cdot  &\cdot &\cdot      &&\cdot  &\cdot &\la^2          &&\cdot  &\sqrt{\frac{t}{t+1}} &\cdot \\
\\
\cdot &\cdot &\sqrt{\frac{t}{t+1}}             &&\cdot  &\cdot  &\cdot   &&\la^2 &\cdot &\cdot \\
\cdot  &\cdot  &\cdot  &&\cdot &\cdot &\sqrt{\frac{t}{t+1}}         &&\cdot &\mu^2&\cdot \\
1 &\cdot &\cdot    &&\cdot &1 &\cdot           &&\cdot &\cdot &\frac{2}{t+1} \\
\end{array}
\right)
\]
and
\[
X^\tau= \left(
\begin{array}{ccccccccccc}
t      &\cdot      &\cdot       &&\cdot   &\cdot   &\cdot  &&\cdot   &\cdot   &\sqrt{\frac{t}{t+1}}\\
\cdot   &\la^2   &\cdot  &&1     &\cdot     &\cdot    &&\cdot    &\cdot    &\cdot   \\
\cdot   &\cdot   &\mu^2  &&\cdot    &\cdot    &\cdot   &&1    &\cdot      &\cdot \\
\\
\cdot  &1 &\cdot               &&\mu^2   &\cdot  &\cdot    &&\cdot  &\cdot &\cdot  \\
\cdot  &\cdot  &\cdot     &&\cdot  &t &\cdot            &&\cdot  &\cdot &\sqrt{\frac{t}{t+1}} \\
\cdot  &\cdot &\cdot      &&\cdot  &\cdot &\la^2          &&\cdot  &1 &\cdot \\
\\
\cdot &\cdot &1               &&\cdot  &\cdot  &\cdot   &&\la^2 &\cdot &\cdot \\
\cdot  &\cdot  &\cdot  &&\cdot &\cdot &1         &&\cdot &\mu^2&\cdot \\
\sqrt{\frac{t}{t+1}} &\cdot &\cdot    &&\cdot &\sqrt{\frac{t}{t+1}} &\cdot           &&\cdot &\cdot &\frac{2}{t+1} \\
\end{array}
\right).
\]
First of all, two matrices
\[
\left (
\begin{array}{ccc}
t & 1 & 1\\
1 & t & 1\\
1 & 1 & \frac 2{t+1}
\end{array}
\right )
\qquad
\left (
\begin{array}{cc}
\la^2 & \sqrt{\frac{t}{t+1}}\\
\sqrt{\frac{t}{t+1}} & \mu^2
\end{array}
\right ),\qquad t>1
\]
are positive semi-definite with rank two. It follows that
$X$ belongs to $\mathbb T$. We note that $\mathcal R X$ is an
$8$-dimensional space spanned by
\begin{equation}\label{eq:dim_rx}
te_{11}+e_{22}+e_{33},\ e_{11}+te_{22}+e_{33},\
e_{12},\ e_{21},\ e_{23},\ e_{32},\ e_{31},\ e_{13}
\end{equation}
and $\mathcal R X^{\tau}$ is a $5$-dimensional space spanned by
\[
te_{11}+\sqrt{\frac{t}{t+1}}\, e_{33},\
te_{22}+\sqrt{\frac{t}{t+1}}\, e_{33},\ \la e_{12}+\mu e_{21},\
\la e_{23}+\mu e_{32},\ \la e_{31}+\mu e_{13}.
\]

Now, we proceed to show that $X$ is an edge state. It is easy to see
that $\mathcal RX^{\tau}$ has following six rank one matrices
\[
\begin{array}{ccc}
{ \left (
\begin{array}{ccc}
(t^2+t)^{\frac 14} & \cdot & \mu\\
\cdot & \cdot & \cdot\\
\la & \cdot & (t^2+t)^{-\frac 14}
\end{array}
\right ) } & {\left (
\begin{array}{ccc}
\cdot & \cdot & \cdot\\
\cdot & (t^2+t)^{\frac 14} & \la \\
\cdot & \mu & (t^2+t)^{-\frac 14}
\end{array}
\right ) } & {\left (
\begin{array}{ccc}
{\mbox{\rm i}} & \la & \cdot\\
\mu & -{\mbox {\rm i}} & \cdot \\
\cdot & \cdot & \cdot
\end{array}
\right ) }
\\
{\left (
\begin{array}{ccc}
-(t^2+t)^{\frac 14} & \cdot & \mu\\
\cdot & \cdot & \cdot\\
\la & \cdot & -(t^2+t)^{-\frac 14}
\end{array}
\right ) } & {\left (
\begin{array}{ccc}
\cdot & \cdot & \cdot\\
\cdot & -(t^2+t)^{\frac 14} & \la \\
\cdot & \mu & -(t^2+t)^{-\frac 14}
\end{array}
\right ) } & {\left (
\begin{array}{ccc}
-{\mbox{\rm i}} & \la & \cdot\\
\mu & {\mbox{\rm i}} & \cdot \\
\cdot & \cdot & \cdot
\end{array}
\right ) }
\end{array}
\]
up to scalar multiplication.
We note that four matrices in the above list have real entries.
If a rank one matrix $xy^*\in \mathcal R X^{\tau}$ is one of them
then $xy^*=\ol xy^*$.  If $xy^*\in
\mathcal R X^{\tau}$ is one of the following matrices
\[
{\mbox{\rm i}}e_{11}-{\mbox{\rm i}}e_{22}+\la e_{12}+\mu e_{21},\quad
-{\mbox{\rm i}}e_{11}+{\mbox{\rm i}}e_{22}+\la e_{12}+\mu e_{21},
\]
with complex entries,
then $\ol xy^*$ should be
\[
{\mbox{\rm i}}e_{11}+{\mbox{\rm i}}e_{22}+\la
e_{12}-\mu e_{21},\quad -{\mbox{\rm i}}e_{11}-{\mbox{\rm i}}e_{22}+\la e_{12}-\mu e_{21},
\]
respectively. In both cases, we can show that $\ol xy^*$ does not
belong to $ \mathcal R X$ which is spanned by matrices in
\ref{eq:dim_rx}. Consequently, there is no rank one matrix
$xy^*\in \mathcal R X^{\tau}$ with $\ol xy^*\in \mathcal R X$.
This gives us a two parameter family of $(8,5)$ edge states.

\end{document}